\documentclass{article}

\begin{document}

\title{Physical Fields in QED}
\author{Othmar Steinmann\\Fakult\"at f\"ur Physik\\Universit\"at
Bielefeld\\D-33501 Bielefeld, Germany}
\date{}
\maketitle
\vspace{-6mm}
\begin{center}
{\em Dedicated to Jacques Bros, an  esteemed colleague and friend}

\end{center}

\vspace{2mm}

\begin{abstract}
The connection between the Gupta-Bleuler formulation and the
Cou\-lomb gauge formulation of QED is discussed. It is argued that
the two formulations  are not connected by a gauge transformation.
Nor can the state space of the Coulomb gauge be identified with a
subspace of the Gupta-Bleuler space. Instead a more indirect
connection between the two formulations via a detour through  the
Wightman reconstruction theorem is proposed.
\end{abstract}

\section{Introduction}
This article is concerned with a major unsolved problem of QED,
that of an exact formulation of the notion of gauge invariance,
more especially the problem of an exact characterization of gauge
transformations and their uses. Why are such seemingly disparate
formulations as the Gupta-Bleuler (GB) formalism and the Coulomb
gauge (C gauge) description physically equivalent, and how are
they connected? Needless to say, this problem will not be solved
here or even fully described. I will merely put forward a few
possibly useful remarks and suggestions. Attention will be
restricted to the two most widely used `gauges' already mentioned,
the GB and the C gauges. And since there still does not exist a
rigorous formulation of QED in any gauge, these results will be
based on the experience gained  in perturbation theory (PT). Any
result which is valid in every order of PT has a good chance of
describing a feature present in a possibly existing exact theory.
That this statement must be taken with a grain of salt will become
apparent later on.

For the structural studies we have in mind, the Wightman functions
are more convenient tools than the Green's functions of the
traditional formulations. The PT of the Wightman functions of QED
has been developed in \cite{1}. Our results are based on the rules
derived there. But the reader's acquaintance with these rules will
not be assumed. The claims made will therefore in general be
substantiated by somewhat heuristic arguments rather than full
proofs. The emphasis is on statements of facts and the discussion
of their significance, rather than on proofs. However, all the
results claimed {\em can} be rigorously proved to all orders of
PT. And they are usually plausible enough as they stand.

\section{Formal Considerations}

The basic fields of QED with charged particles of spin 1/2 are the
electromagnetic potentials $A_{\mu}(x)$ and the Dirac spinors
$\psi(x)$ and $\bar{\psi}(x)=\psi^*(x)\gamma^0$.

The {\em GB formalism} has the advantage of working with local,
covariant, fields $A_{\mu},\,\psi,$ as we like them. But it also
has two grave drawbacks. First, its state space ${\cal V}_{GB}$ is
equipped with an indefinite scalar product, hence it is not a
Hilbert space. This contradicts the basic rules of quantum
mechanics, and it is mathematically inconvenient. Second, the
Maxwell equations

\begin{equation}
\partial_{\nu} F^{\nu\mu}(x) = j^{\mu}(x)\ ,         \label{1}
\end{equation}
with

\begin{equation}
F^{\nu\mu}(x) =
\partial^{\nu}A^{\mu}(x)-\partial^{\mu}A^{\nu}(x)\,,\qquad
j^{\mu}(x) = e\,\bar{\psi}(x)\gamma^{\mu}\psi(x)\ ,   \label{2}
\end{equation}
are not satisfied as operator equations on ${\cal V}_{GB}$, even
after renormalization. This means that ${\cal V}_{GB}$ contains
unphysical states, that is vectors which do not describe states
ever encountered in a laboratory. This raises the question of how
to characterize the physical states, and the second question
whether ${\cal V}_{GB}$ can be defined such that it contains
sufficiently many physical states to give a full description of
reality. The standard textbook answer to the first question is
this: the divergence
\begin{equation}
B(x) = \partial_{\mu}A^{\mu}(x)                      \label{3}
\end{equation}
solves the free field equation $\partial^{\mu}\partial_{\mu}
B(x)=0$, hence can be split into a creation part $B^-(x)$ and an
annihilation part $B^+(x)$. And the physical subspace  ${\cal
V}_{ph}\subset{\cal V}_{GB}$ is defined to be the kernel of $B^+$:
\begin{equation}
B^+(x)\,{\cal V}_{ph} = 0\,.                     \label{4}
\end{equation}
This condition ensures the validity of the Maxwell equations on
${\cal V}_{ph}$. The second question is, as a rule, simply
ignored. A clean definition of ${\cal V}_{GB}$  and thus of
${\cal{V}}_{ph}$ is hardly ever given.\footnote{There exist more
thoughtful treatments of the problem outside of textbooks, for
example \cite{2,3,4}. These approaches make essential use of the
notion of asymptotic fields, which is not unproblematic in QED and
will be avoided in the present work.}

The {\em C gauge} has complementary advantages and drawbacks. Its
state space ${\cal V}_C$ is a Hilbert space and contains only
physical states. The Maxwell equations are satisfied. But the
basic fields $A_{\mu},\:\psi,\:\bar{\psi},$ are neither local
(i.e. they do not commute at spacelike distances) nor covariant.
This is very inconvenient for a detailed {\em ab ovo} elaboration
of the theory, especially for a convincing formulation of
renormalization. And the lack of manifest covariance also raises
the question of how observational Lorentz invariance emerges from
the theory.

The complementary aspects of the two methods suggests a joining of
their forces. This may consist in first working out the theory of
the GB-fields $A_{\mu},\:\psi,$ as fully as possible, and then
identifying in this framework ``physical fields" ${\cal
A}_{\mu},\:\Psi,$ which generate ${\cal V}_{ph}$  out of the
vacuum $\Omega$, yielding the C gauge formulation. Formally, such
``C-fields" are obtained from the GB-fields by the Dirac ansatz
\begin{eqnarray}
{\cal A}_{\mu}(x) & = & A_{\mu}(x) - \frac{\partial}{\partial
x^{\mu}} \int dy\,r^{\rho}(x-y)\,A_{\rho}(y)\,,  \nonumber  \\
\Psi(x) & = & \exp\Big\{ie\int
dy\,r^{\rho}(x-y)\,A_{\rho}(y)\Big\}\,\psi(x)\,,    \label{5}  \\
\bar{\Psi}(x) & = & \Psi^*(x)\,\gamma^0\,,   \nonumber
\end{eqnarray}
with
\begin{equation}
r^0=0\,,\qquad r_j(x)=-r^j(x)=\delta(x^0)\,\partial_j
\frac{1}{|\vec{x}|}\, .             \label{6}
\end{equation}
for $j=1,2,3$. The auxiliary `functions' $r^{\mu}$ satisfy
\begin{equation}
\partial_{\mu} r^{\mu}(x)  = \delta^4(x)\ ,     \label{7}
\end{equation}
or in momentum space
\begin{equation}
   p_{\mu}\tilde{r}^{\mu}(p) = i             \label{8}
\end{equation}
with $\tilde{r}_{\mu}(p)=\int dx\,\exp\{ipx\}\,r^{\mu}(x)$. Notice
that {\em formally} the definitions (\ref{5}) have the form of a
gauge transformation with the operator valued and field-dependent
gauge function
\begin{equation}
G(x) = - \int dy\,r^{\rho}(x-y)\,A_{\rho}(y)\;.   \label{9}
\end{equation}
That the fields (\ref{5}) really are the desired C-fields follows
from the fact that they satisfy
\begin{equation}
\big[B(x),\,\Psi(y)\big] = \big[B(x),\,\bar{\Psi}(y)\big] =
\big[B(x),\,{\cal A}_{\mu}(y)\big] = 0\;,        \label{10}
\end{equation}
which equations then also hold for the annihilation part $B^+$.
And this together with $B^+\Omega=0$ shows that states of the form
\begin{equation}
\Phi = {\cal P}(\Psi,\,\bar{\Psi},\,{\cal A}_{\mu})\,\Omega\;,
\label{11}
\end{equation}
with ${\cal P}$ a polynomial -- or a more general function, if
definable -- of fields averaged over test functions, are physical
in the sense of (\ref{4}). The property (\ref{10}) has been called
``strict gauge invariance" by Symanzik \cite{5} and, following
him, by Strocchi and Wightman \cite{6}. A proof of (\ref{10})  is
found in the first of these references.

\section{Problems, and a Solution}

The contents of Sect.~2 were purely formal. In order to give the
definition (\ref{5}) a rigorous meaning we must start from a
rigorous description of the GB formalism, especially of ${\cal
V}_{GB}$. The use of a suitably adapted version of the Wightman
formalism \cite{7} suggests itself. Of course, the naturally
defined scalar product being indefinite, ${\cal V}_{GB}$ cannot be
a Hilbert space. But we assume it to be equipped with a
non-degenerate scalar product. And we wish to retain all the other
Wightman axioms. In particular, the fields
$A_{\mu},\,\psi,\,\bar{\psi},$ are to be operator valued tempered
distributions. And we assume the subspace ${\cal V}_0$ spanned by
tempered field polynomials applied to the vacuum $\Omega$ to be
dense in ${\cal V}_{GB}$ in the weak topology induced by the
scalar product. ${\cal V}_0$ is called the space of ``local
states". The density of ${\cal V}_0$ is important because it
allows to construct ${\cal V}_{GB}$ from the Wightman functions by
the reconstruction theorem. The Wightman functions are easily
calculated in PT and other schemes of approximation. And their
properties, as properties of tempered distributions, are easier to
investigate than those of unbounded operators in a space with
indefinite metric. On ${\cal V}_0$ the Wightman functions
determine the scalar product.

Right at the start we are confronted with a slightly disturbing
fact. Let the charge operator $Q$ be defined by
\begin{equation}
[Q,\,\psi(x)]=-e\,\psi(x)\,,\quad
[Q,\,\bar{\psi}(x)]=e\,\bar{\psi}(x)\,, \quad [Q,\,{\cal
A}_{\mu}(x)] = 0\,,\quad Q\,\Omega = 0\;,               \label{12}
\end{equation}
with $e$ the charge of the positron. Define the state $\Phi$ to be
{\em physical} if the Gau\ss\  law
\begin{equation}
Q\,\Phi = \int_{x^0=t}d^3x\,\nabla\vec{E}(x)\,\Phi    \label{13}
\end{equation}
holds on it\footnote{As is well known, in this crude version of
the condition the right-hand side does not make sense. Suitable
regularizations in space and time  are necessary. But this problem
is immaterial to our present purposes.}, with $\vec{E}$ the
electric field strength.  Then it is known \cite{8} that ${\cal
V}_0$ contains no charged physical states. This means that the
construction of charged physical states in ${\cal V}_{GB}$ as
limits of local states is a non-trivial task.\footnote{That it is
not necessarily an unsolvable task has recently been
re-em\-pha\-sized by Morchio and Strocchi \cite{8a}.}

Let us now turn to the problem of giving a rigorous meaning to the
equations (\ref{5}) defining the C-fields in terms of the
GB-fields, supposing a rigorous theory of the latter to be at hand
as just explained. In this attempt we encounter two problems, an
ultraviolet (UV) one and an infrared (IR) one.

The UV problem is this: The factor $\psi(x)$ in $\Psi(x)$ is a
distribution, not a function, and so is the exponential factor.
Notice that the auxiliary `functions' $r^j$ are not functions in
the strict sense of the word, let alone test functions, so that
the exponent does not exist as a function. This problem can be
solved by standard renormalization procedures, most easily by
subtraction at $p=0$ in momentum space (`intermediate
renormalization'). Unfortunately such a subtraction destroys the
product form of $\Psi$, because in $p$-space the product
$a(x)b(x)$ becomes the convolution $\int dk\,\tilde{a}(p-k)\,
\tilde{b}(k)$ (the tilde denotes the Fourier transform). This
reads in its subtracted form
\[ \int dk\,\big[\tilde{a}(p-k)\,\tilde{b}(k)-\tilde{a}(-k)\,
 \tilde{b}(k)\big]\, ,   \]
which is no longer a convolution. Hence the mapping
$(\psi,A)\to(\Psi,{\cal A})$ has no longer the form of a gauge
transformation. It is very unlikely that this problem can be
solved by a more sophisticated  method of renormalization. {\em It
is my opinion that we should not bemoan this fact, but accustom
ourselves to the idea that the notion of gauge transformations is
not as useful in QED as it is in classical electrodynamics, but is
 only of a heuristic value}.\footnote{This scepticism does not extend
 to gauge transformations of the first kind, that is global
  transformations with an $x$-independent real gauge function.} The true
  problem is to find fields satisfying the condition (\ref{10})  of
  strict gauge invariance, no matter whether or not they can be
  derived from the GB-fields by a gauge transformation. That the
  renormalized Dirac ansatz yields a formulation of QED which
  satisfies all the necessary requirements, fully justifies its
  use.

The IR problem is this: Are $\Psi,\,{\cal A}_{\mu}$, after
renormalization, defined as fields on ${\cal V}_{GB}$? In other
words, is the space ${\cal V}_C$ generated from the vacuum
$\Omega$ by the C-fields a subspace of ${\cal V}_{GB}$? At first,
the answer gleaned from PT is a plain `no'! The scalar product
$(\Phi_0,\,\Phi)$ of the physical state
\begin{equation}
\Phi = \int dx\,f(x)\,\Psi(x)\,\Omega\quad \in{\cal V}_C
\label{14}
\end{equation}
and the local state
\begin{equation}
\Phi_0 = \int dy\,\overline{g(y)}\,{\bar{\psi}}^{\,*}\,\Omega
\quad \in{\cal V}_0\;,                     \label{15}
\end{equation}
with $f$ and $g$ tempered test functions, can be shown to diverge
already in second order of PT (see end of Appendix). This
divergence is caused by the divergence at large $y$ of the
exponent $\int dy\,r^{\rho}(x-y)\,A_{\rho}(y)$ in (\ref{5}): it is
an IR problem.

But this result is misleading. As is well known, the generic
$S$-matrix element of QED calculated with the LSZ reduction
formula is in general IR divergent in finite orders of PT. But
these divergences can be isolated and summed over all orders,
yielding a {\em vanishing} result. And this result is expected to
be closer to the truth than the finite-order divergences, because
it agrees with information obtained from other sources, especially
the Bloch-Nordsieck model. Something similar might happen for the
mixed 2-point function
\begin{equation}
F(x,y) = (\Omega,\,\bar{\psi}(y)\,\Psi(x)\,\Omega) \label{16}
\end{equation}
occurring in $(\Phi_0,\,\Phi)$. And, indeed, it does happen! The
IR divergences in $F$ can be isolated in all orders of PT and
summed to yield
\begin{equation}
F(x,y) \equiv 0\;.              \label{17}
\end{equation}
For a sketch of the proof we refer to the Appendix. Again, we
expect the result (\ref{17}), rather than the finite-order
divergences, to correspond to the true situation. But, as in the
case of the $S$-matrix, this does not solve our problem. In the
same way as (\ref{17}) it can be generally shown that $\Phi$ is
orthogonal to all local states. The state $\Phi$ of charge $-e$ is
orthogonal to ${\cal V}_0$, hence it cannot be the weak limit of a
sequence of local states, hence ${\cal V}_C$
 cannot be a subspace of ${\cal V}_{GB}$, in which we assumed ${\cal
 V}_0$ to be dense.

As a result we obtain:

\medskip

\begin{it}
The state space ${\cal V}_C$ of the Coulomb gauge cannot be
obtained as a subspace of an extension ${\cal V}_{GB}$ of ${\cal
V}_0$, if the scalar product defined on ${\cal V}_0$ can be
extended to ${\cal V}_{GB}$  in such a way that ${\cal V}_0$ is
weakly dense in ${\cal V}_{GB}$.
\end{it}

\medskip

The problems discussed in this section lead to the following
conclusions. The C-fields $\Psi,\,{\cal A}_{\mu}$, cannot be
defined as fields acting on ${\cal V}_{GB}$. Nor are they related
to the GB-fields $\psi,\,A_{\mu}$, by a gauge transformation. The
traditional explanations of the connection between the two
formulations do not work. A working method of relating the two
formalisms has been derived in Chap.~12 of \cite{1}. It will be
briefly described without giving proofs. The method makes
essential use of the Wightman reconstruction theorem. At first it
is demonstrated (in PT) that the Wightman functions of the
C-fields can be obtained from those of the GB-fields by a limiting
procedure, like this: Replace the auxiliary functions $r^j(x)$ in
(\ref{5}) and (\ref{6}) by the regularized version
\begin{equation}
r^j_{\xi}(x) = \chi(\xi x)\, r^j(x)\;,\qquad 0<\xi<\infty\;,
\label{18}
\end{equation}
with $\chi(u)$ a test function with compact support, satisfying
$\chi(0)=1$. The resulting fields $\Psi_{\xi},\,{\cal
A}^{\mu}_{\xi}$ {\em are} definable as acting on a slight
extension of ${\cal V}_0$. Their Wightman functions can be
computed. And the limits $\xi \to 0$ of the latter exist and
define a field theory via the reconstruction theorem, which has
all the desired properties of QED in the C gauge. It {\em is} QED
in the C gauge! But the states and the fields of this theory have
no discernible direct connection with the states and fields of the
GB formalism.

\medskip

Let us end this section with a few remarks on how observational
Lorentz invariance emerges from the not manifestly covariant C
gauge formalism. The problem has not yet been discussed in depth.
I can therefore only state a program rather than results. This
program is based on the spirit of the local observables approach
\cite{9} to quantum field theory, according to which only
observables localized in bounded regions of space-time are true
observables. The first problem facing us is finding an exact
characterization of the operators representing observables. Since
the C-fields should describe the theory fully, an observable in
the bounded domain ${\cal R}$ must be a function of the fields
with arguments in ${\cal R}$. Observables should be represented by
hermitian operators (self-adjointness is hard to handle in PT),
and they should satisfy the axioms of local quantum physics. In
particular, two observables localized in relatively spacelike
domains should commute. And the algebras of observables in the
domain ${\cal R}$ and its image ${\cal R}_{\Lambda}$ under the
Lorentz transformation $\Lambda$ should be related by an
automorphism $\alpha_{\Lambda}$, which we can, of course, not
expect to be unitarily implemented. The traditional requirement
that observables should be gauge invariant is difficult or
impossible to formulate in C gauge, on account of the doubtful
status of gauge transformations. We propose to solve this
conundrum like for the fields, by starting from the GB formalism.
There, observables must satisfy the same requirements as stated
for the C gauge. But the additional requirement of gauge
invariance can now be interpreted to denote strict gauge
invariance. This means that observables must commute with
$B(x)$.\footnote{On ${\cal V}_C$ $B$ vanishes identically.} The C
gauge equivalent of such a GB observable $A$ can then be defined
as a bilinear form by
\begin{equation}
(\Omega,\,\bar{\Psi}(x)\,A\,\Psi(y)\,\Omega) = \lim_{\xi\to0}
(\Omega,\,\bar{\Psi}_{\xi}(x)\,A\,\Psi_{\xi}(y)\,\Omega)
\label{19}
\end{equation}
and obvious generalizations. Under exactly what conditions on $A$
this limit exists has not yet been investigated. It is clear that
the observable fields $F_{\mu\nu}$ and $j_{\mu}$ belong to this
class.

Given the local observables of the theory, the second problem is
that of describing observational Lorentz invariance without using
a non-existent unitary representation of the Lorentz group. The
formulation proposed here is based on the important insights put
forward by Haag and Kastler in \cite{10} concerning the relevance
of Fell's theorem (a purely mathematical statement) to quantum
mechanics. This relevance rests on the observation that in any
given experiment we can only measure a {\em finite number} of {\em
local} observables with a {\em finite accuracy}. In view of this
we can formulate observational invariance as follows.

\medskip

\begin{it}
Let $A_1,\cdots,A_n,$ be a finite set of local observables with
measuring accuracies $\epsilon_i$,
$A_i^{\Lambda}=\alpha_{\Lambda}(A_i)$  their images under the
Lorentz transformation $\Lambda$, and let $\Phi$ be a (physically
preparable) state in ${\cal V}_C$. Then there exists a state
$\Phi_{\Lambda}\in{\cal V}_C$ such that
\begin{equation}
|(\Phi_{\Lambda},\,A_i^{\Lambda}\,\Phi_{\Lambda}) -
(\Phi,\,A_i\,\Phi)|<\epsilon_i        \label{21}
\end{equation}
for $i=1,\cdots,n$.
\end{it}

\medskip

If $\Lambda$ is a boost, then $\Phi_{\Lambda}$ represents the
original state as seen by a moving observer,  as far as the
experiment in question is concerned. That this form of Lorentz
invariance is satisfied in ${\cal V}_C$ is at the moment still a
conjecture. But it has a good chance of being true. Partial
results in this direction can be found at the end of Chap.~12 in
\cite{1}.

\section*{Appendix: {\large Proving Equation (\ref{17})}}
\setcounter{equation}{0}
\renewcommand{\theequation}{A.\arabic{equation}}

Using the methods of \cite{1} a rigorous proof of the claimed
result \ref{17} can be given  by summing the IR relevant parts of
$F$ in all orders of PT. In this appendix we will not give the
full proof, but only sketch its essential ideas.

We denote the vacuum expectation value $(\Omega,\cdots \Omega)$ by
$\langle\cdots\rangle$. The expression of interest is
\begin{eqnarray}
A & = & \langle \bar{\psi}(y)\,\Psi(x)\rangle    \nonumber \\
   & = & \langle\bar{\psi}(y)\,\exp\{ie\int du\,r^j(x-u)\,
   A_j(u)\} \,\psi(x)\rangle       \label{A1}
\end{eqnarray}
with
\begin{equation}
r^j(v) = \delta(v^0)\,\partial^j \frac{1}{|\vec{v}|}\,. \label{A2}
\end{equation}
We are only interested in the IR problems connected with this
expression, ignoring the UV problems which can, however, easily be
taken into account. The reader is free to get rid of them by a
suitable UV regularization. This means that our problem is the
existence of the $u$-integral in (\ref{A1}) {\em at large} $u$.
Assume $x,y,$ to be restricted to bounded regions by integrating
them over test functions  with compact support. Again, this
restriction is not essential. Then, because of the $\delta$-factor
in (\ref{A2}), $u$ can tend to infinity only in spacelike
directions. For large $u$ we can then neglect $\vec{x}$ with
respect to $\vec{u}$ in $r^j(x-u)$. Using the cluster property we
find
\begin{equation}
A \sim \langle\psi(x)\,\bar{\psi}(y)\rangle\langle \exp\{ie\int
du\,r^j(-u)\,A_j(u)\}\rangle\;,       \label{A3}
\end{equation}
where the symbol $\sim$ means that only the potentially IR
dangerous contributions to $A$ are considered. The first factor
depending on $x$ and $y$ is IR harmless. We need therefore only
discuss the second factor, which we call $X$. Going over to
momentum space we find
\begin{equation}
X = \Big\langle\exp\Big\{e\int
dk^0\,d^3k\,\frac{k^j}{|\vec{k}|^2}\,\tilde{A}_j(k)
\Big\}\Big\rangle \label{A4}
\end{equation}
with $\tilde{A_j}$ the Fourier transform of $A_j$ (the tilde will
be omitted in the sequel). The existence of the $k^0$-integral is
part of the UV problem which we ignore. The same goes for for the
existence of the $\vec{k}$-integral at $|\vec{k}|\to\infty$. Our
only concern is the possible divergence of the $\vec{k}$-integral
at $\vec{k}=0$. In PT the exponential in (\ref{A4}) is defined as
a power series. (Notice the factor $e$ in the exponent.) Only the
even terms in this expansion survive because the Wightman
functions of an odd number of $A $'s and no $\psi$ or $\bar{\psi}$
vanish. Hence we find
\begin{equation}
X = \sum^{\infty}_{\sigma=0}\frac{e^{2\sigma}}{(2\sigma)!}
\bigg\langle\Big(\int dk\,\frac{k^j}{|\vec{k}|^2}\,
A_j(k)\Big)^{2\sigma}\bigg\rangle\;.         \label{A5}
\end{equation}
This expression contains terms of the form
\[  \left\langle\prod^{2{\sigma}}_{\omega=1}
A_{j_{\omega}}(k_{\omega})\right\rangle \] which must be evaluated
in PT, and its IR divergences isolated. The graph rules for these
functions are similar to, but somewhat more complicated than, the
familiar Feynman rules for Green's functions. As in this case it
is found that connected components with $\ge4$ external lines of a
graph give convergent contributions to $X$, because they vanish
sufficiently strongly at $k_{\omega}=0$ to overcome the
$1/|\vec{k_{\omega}}|$ singularities in (\ref{A5}). This is a
consequence of the Ward identities,  which for these fermion-less
connected graphs take the form $k^{\mu}A_{\mu}(k)=0$ in all orders
except the $0^{th}$. These contributions can be factored out. The
IR problem is concentrated in graphs consisting exclusively of
2-variable components. But such a 2-point function is assumed to
satisfy the renormalization condition
\begin{equation}
\langle A_{\mu}(k_1)\,A_{\nu}(k_2)\rangle = \delta^4(k_1+k_2)\,
\big(g_{\mu\nu} \delta_+(k_1) + {\rm IR\ harmless\ terms}\big)
\label{A6}
\end{equation}
with $\delta_+(k)=\theta(k_0)\,\delta(k^2)$. The IR divergences
are entirely due to the zero-order term $g_{\mu\nu}\delta_+$.
Inserting this into the expansion (\ref{A5}) and doing the
combinatorics right we find
\begin{equation}
X \sim {\rm (finite\ factor)}\:\cdot\:\exp\Big\{-e^2 \int
\frac{d^3k}{2|\vec{k}|^3}\Big\}\;.        \label{A7}
\end{equation}
The $k$-integral diverges positively at $\vec{k}=0$. The UV
divergence at $|\vec{k}|\to\infty$ is in a more detailed treatment
removed by renormalization and does not concern us here. The
result is, then:
\begin{equation}
X \sim 0                    \label{A8}
\end{equation}
which proves (\ref{17}). Notice that the $e^2$ term in the power
series expansion of (\ref{A7}) diverges, as was claimed in
Sect.~3.


\begin{thebibliography}{11}
\bibitem{1} O.~Steinmann: {\em Perturbative QED and Axiomatic Field
Theory}. Berlin, Springer 2000.
\bibitem{2} D.~Zwanziger: Phys.~Rev.~D14, 2570 (1976).
\bibitem{3} J.~Fr\"ohlich, G.~Morchio, and F.~Strocchi: Ann.~of
Phys.~119, 2441 (1979).
\bibitem{4} G.~Morchio and F.~Strocchi: Nucl.~Phys.~B211, 471
(1983).
\bibitem{5} K.~Symanzik: {\em Lectures on Lagrangian Quantum Field
Theory}. Internal report DESY T71-1 (1971).
\bibitem{6} F.~Strocchi and A.~S.~Wightman: J.~Math.~Phys.~15, 21
(1974).
\bibitem{7} R.~F.~Streater and A.~S.~Wightman:
{\em PCT, Spin \& Statistics, and All That}, $2{nd}$ edition.
Reading MA, Benjamin/Cummings 1978. -- R.~Jost: {\em The General
Theory of Quantized Fields}. Providence RI, Am.~Math.~Soc.~1965.
-- N.~N.~Bogolubov, A.~A.~Logunov, A.~I.~Oksak, and I.~T.~Todorov:
{\em General Principles of Quantum Field Theory.} Dordrecht,
Kluwer 1990.
\bibitem{8} R.~Ferrari, L.~Picasso, and F.~Strocchi:
Commun.~Math.~Phys.~35, 25 (1974).
\bibitem{8a} G.~Morchio and F.~Strocchi: J.~Math.~Phys.~44, 5569
(2003).
\bibitem{9} R.~Haag: {\em Local Quantum Physics.} Berlin, Springer
1993.
\bibitem{10} R.~Haag and D.~Kastler: J.~Math.~Phys.~5, 848 (1964).
\end{thebibliography}
\end{document}